\documentstyle[prl,twocolumn,aps,epsf]{revtex}

\begin{document}
\draft
\date{May 3, 1996}
\title{Electron Wave Filters from Inverse Scattering Theory}
\author{Daniel Bessis\cite{db}, Giorgio Mantica\cite{gm}, 
G. Andrei Mezincescu\cite{gam} and Daniel Vr\^\i nceanu\cite{dv}}
\address{Center for Theoretical Study of Physical Systems, 
Clark - Atlanta University, Atlanta, GA 30314}
\maketitle

\begin{abstract}
Semiconductor heterostructures with prescribed energy dependence of
the transmittance can be designed by combining: {\em a)} Pad\'e approximant
reconstruction of the S-matrix; {\em b)} inverse scattering theory for
Schro\"dinger's equation; {\em c)} a unitary transformation which takes into
account the variable mass effects. The resultant continuous concentration
profile can be digitized into an easily realizable rectangular-wells
structure. For illustration, we give the specifications of a 2 narrow band-pass
12 layer $Al_cGa_{1-c}As$ filter with the high energy peak more than {\em
twice narrower} than the other.
\end{abstract}


\narrowtext
The inverse scattering method \cite{KM55,Sab83a,CS89} 
allows the reconstruction of the potential, $V(x)$, in the
one dimensional Schr\"odinger equation,  
\begin{equation}
-\frac{\hbar^2}{2m_0}\psi^{\prime\prime}+V(x)\psi =E\psi, 
\label{1}\end{equation}
from the energy dependence of the S-matrix (complex reflection 
and transmission coefficients). Let us briefly review 
the necessary results, referring to \cite{CS89} for details.
Setting $E=\hbar^2k^2/2m_0>0$, the transmission, $T(k)$,  and 
the left/right reflection coefficients, $R_\pm(k)$ satisfy 
\begin{eqnarray}
R_+(k)T(-k)+R_-(-k)T(k)=0,\label{4}\\
T(k)T(-k)+R_\pm(k)R_\pm(-k)=1,\label{5}\\
T(k)=\left[T(-k)\right]^*;   ~~
R_\pm(k)=\left[R_\pm(-k)\right]^*, \label{6}
\end{eqnarray}
where ~$^*$~ denotes complex conjugation. 
In addition, $|R_\pm(k)|<1$, except at 
$k=0$, where generically one has $R_-(0)=-1$. Moreover, 
$\lim_{k\to\infty}R_\pm(k)=0$.

The scattering coefficients $T(k)$ and $R_\pm(k)$ 
can be continued analytically to the upper 
half-plane of the complex variable $k$. 
We will consider only the case when  
$V(x)=0$ for $x<0$ and there are no bound states. 
Then, for ${\hbox{\rm Im}}(k) > 0$, $T(k)$ 
and the left-reflection coefficient, $R_-(k)$, 
are analytic functions having no zeros.
The right-reflection coefficient, $R_+(k)$, 
may have additional poles. The relations 
(\ref{4},\ref{5}) remain valid in the whole upper 
complex half-plane, while eqs. (\ref{6})
hold with $-k$ replaced by $-k^*$ in the right 
hand sides. 

The potential $V(x)$ is reconstructed as 
\begin{equation}
V(x)=\frac{\hbar^2}{m_0}\frac{{\rm d}}{{\rm d}x}K_-(x,x), 
\end{equation}
where the function $K_-(x,y)$ satisfies Marchenko's 
integral equation
\begin{equation}
K_-(x,y)+M_-(x+y)+\int_{-y}^xM_-(y+s)K_-(x,s){\rm d}s=0.
\label{marce}
\end{equation}
Here, $M_-(u)$ is the Fourier transform of the (complex) 
left-reflection coefficient, $R_-(k)$, which is therefore 
the input information required for the inverse technique.  

Although inverse scattering methods have been 
extensively used in various branches of physics 
and engineering \cite{CS89,etc}, they have not 
yet been used in heterostructure design. 
In this Letter we give a brief outline of the application 
of these techniques to the design of 
electron-wave filters --- semiconductor heterostructures 
having a prescribed transmittance, $|T(k)|^2$. So far,
such filters had been designed using methods adapted from 
optical filter synthesis \cite{GGB89}. 
These structures may be useful for far-infrared lasers and 
detectors \cite{BK89,Capasso}. 

In the effective mass -- envelope function approximation, 
the electron states in the conduction band of a heterostructure 
having a position dependent composition ($x$ is the
deposition direction and $c(x)$ the alloy concentration)
are solutions of the BenDaniel and Duke equation 
\cite{BDD66,Bas88,GK93a}:
\begin{equation}
-\nabla\frac{\hbar^2}{2m(x)}\nabla\psi + {\cal E}_c(x)\psi = 
E\psi, \label{2}
\end{equation}
where $m(x)=m_{eff}[c(x)]$ and 
${\cal E}_c(x)={\cal E}_{cond}[c(x)]$ 
are the local (isotropic) effective mass and conduction band offset
of conduction band electrons. 
We consider here only the case when the device is 
embedded (at least in an asymptotic sense) in a crystal
with uniform composition, {\it i.e.} we assume that 
$c=c_\infty$ for $x<0$ and in the limit for $x\longrightarrow +\infty$; 
let $m_\infty$ be the conduction-band effective mass in the embedding 
material. In many systems of interest, 
such as $Al_cGa_{1-c}As$, the dependence of $m_{eff}$ and 
${\cal E}_{cond}$ on $c$ is approximately linear \cite{something}: 
$m_{eff}(c)=a+bc$, ${\cal E}_{cond}(c)=\alpha +\beta c$. 
The mass - concentration dependence can then be inverted,
to get:
\begin{equation}
{\cal E}_{c}(m)={\cal A + B}m.\label{2a}
\end{equation}
For $Al_cGa_{1-c}As$ we use the numerical values 
${\cal A}=-624.1$meV and ${\cal B}=9315$meV/$m_{e}$, 
where $m_e$ is the free electron mass. 

After separating away the variables associated with motions
orthogonal to the $x$ direction in eq. (\ref{2}) and 
setting the transverse quasimomentum ${\bf q}_\perp = 0$, 
the envelope function for the $x$ direction, $f$, satisfies:
\begin{equation}
-\left(\frac{\hbar^2f^\prime}{2m}\right)^\prime +
{\cal B}\bigl[m(x)-m_\infty\bigr]f 
= \frac{\hbar^2k^2}{2m_\infty}f.
\label{3}\end{equation}
In the above, we have used (\ref{2a}) and set
\begin{equation}
\hbar k=\sqrt{2m_\infty E}, 
\label{3a}\end{equation}
the energy $E$ being measured from the conduction 
band edge of the embedding crystal. An additional term 
$E_t({\bf q}_\perp)(\left[m_\infty/m(x)-1\right]f(x)$, 
with $E$ in (\ref{3a}) replaced by $E-E_t({\bf q}_\perp)$, 
where $E_t({\bf q}_\perp)=(1/2)\hbar^2{\bf q}_\perp^2/2m_\infty$,  
appears in the left hand side of (\ref{3}) for 
${\bf q}_\perp\ne 0$. In $Al_cGa_{1-c}As$ the effect of this 
corrections is quite small for $E_t({\bf q}_\perp)\approx k_BT$, 
up to room temperature. Eq. (\ref{3}) is  similar to the 
one-dimensional Schr\"odinger equation (\ref{1}), but with a 
position-dependent mass. This variability has nontrivial dynamical 
effects \cite{gio}. Noticing that only $m(x)$ enters (\ref{3}), 
our task is now to determine mass (and hence concentration $c(x)$) 
profiles giving rise to the pre-assigned transmissivity $|T(k)|^2$. 
Let us highlight the main problems and our approach to solving them.

1. Eq. (\ref{2}) is a low energy approximation, valid only  for 
the sufficiently small energies at which the conduction band 
energy is parabolic. Since we assume to be interested only in  
this range of energies, the transmittance behavior can  
be arbitrarily continued to higher energies, respecting the 
analyticity properties of the $S$-matrix. The design problem now 
has an infinity of solutions. We will not explore all of them, 
but we will use Pad\'e approximants \cite{pade} to fit the 
transmittance specifications within given tolerances in the 
required energy range. The non-unicity of the approximants will give 
free parameters which can be used in successive developments to 
optimize properties of the device other than the energy 
dependence of $|T|^2$. 

2. The transmittance input data we use give only
$|T(k)|^2$ on the real axis, while the solution
of the inverse problem for (\ref{1}) requires also 
the phases of $R_\pm$. 
Generally speaking, the necessary phases can be 
reconstructed by using dispersion relations as in 
particle theory \cite{dispersion}. Yet, since we use 
Pad\'e (rational) approximants for fitting the energy 
dependence of $|T|^2$, we can perform the analytic 
continuation in a simpler way. In fact, in the case 
when no bound states are present, we let  
\begin{equation}
{\cal R}(E)=\frac{{\cal P}_p(E)}{{\cal Q}_q(E)}, \label{7}
\end{equation}
be the Pad\'e approximant fitted to the reflectance. 
${\cal P}_p(E)$ and ${\cal Q}_q(E)$ are polynomials of degree 
$p$ and $q\ge p+2$ with {\em real} coefficients. The condition 
$q\ge p+2$ guarantees that ${\cal R}(E)={\cal O}(E^{-2})$ 
for large $E$ and, therefore, the potential is non-singular 
at $x=0$. In accordance with the analytic properties of the 
reflectance on the real axis, (\ref{4}-\ref{6}), 
$0\le{\cal R}(E)<1$ for $0<E<\infty$ and ${\cal R}(0) =1$. 
The latter implies that ${\cal P}_p(0)={\cal Q}_q(0)=1$. From 
the former follows that if ${\cal P}_p(E)$ has real positive 
zeros (which correspond to {\it resonances}, 
where the transmittance attains its maximal allowed 
value, unity) their multiplicities have to be even. 
 From the structure of eq. (\ref{marce}), we see that it
is convenient to work in $k$ space, and to compute
$(p,q)$ Pad\'e approximants to the 
left-reflection coefficient  
\begin{equation}
R_-(k)=\frac{P_p(k)}{Q_q(k)}, \label{P1}
\end{equation}
where the $k=0$ values of the polynomials are taken as
$P_p(0)=-Q_q(0)=-1$. This guarantees that $R_-(0)=-1$. 
Let $k_\alpha$ (respectively, $k_j$) be the zeros of $P$ ($Q$). 
To satisfy the analytic properties of $R_-(k)$ and the 
condition (\ref{6}), the zeros of $P$ and $Q$  
must have non-positive imaginary parts (negative for 
$P$). If $k_\alpha$  ($k_j$) is a root of $P(k)=0$ ($Q(k)=0$),  
then $-k_\alpha^*$ ($-k_j^*$) must also be a zero 
of the same polynomial: zeros of $P$ and 
$Q$ which are not purely imaginary come in pairs. 
Using the root product expansion of the polynomials
we may write
$R_-(k)=-\prod_{\alpha=1}^p\left(1-\frac{k}{k_\alpha}\right)/
\prod_{j=1}^p\left(1-\frac{k}{k_j}\right)$. 
For real positive energies the reflectance is given by 
\begin{equation}
{\cal R}(E)=R_-(k)R_-(-k), \label{P4}
\end{equation}
where we used (\ref{6}). Thus, taking into account 
(\ref{3a}), we see that the zeros and poles of 
the analytic continuation to the complex $E$ plane 
of the Pad\'e approximant (\ref{7}) for the 
real axis reflectance are connected to the 
zeros and poles of the Pad\'e approximant 
(\ref{P1}) for the (complex) reflection coefficient 
by $k=\pm\sqrt{2m_\infty E}/\hbar$. The sign assignments 
are {\em unambiguous} in the absence of bound states, when  
$R_-(k)$ is analytic and has no zeros in the upper 
half plane.  The phase reconstruction for the 
analytic continuation of the reflection coefficient 
$R_-(k)$ from the absolute values on the 
real axis is therefore achieved. 

Similarly, we seek the transmission coefficient, 
as a $(q,q)$ Pad\'e approximant having 
the same poles as $R_-(k)$:
\begin{equation}
T(k)=\frac{kL_{q-1}(k)}{Q_q(k)}. \label{T1}
\end{equation}
Here we used $T(0)=0$. The polynomial $L_{q-1}(k)$  
is determined by its roots and the condition 
$\lim_{k\to\infty}T(k)=1$. 
 From (\ref{P4}) and (\ref{5},\ref{6}), the 
transmittance on the real axis is given by 
${\cal T}(E)=1-R_-(k)R_-(-k)$ and (\ref{T1}) leads to 
the following equation for the zeros of $L_{q-1}(k)$  
\begin{equation}
k^2L_{q-1}(k)L_{q-1}(-k)={\cal Q}_q(E)-{\cal P}_p(E)=0. 
\label{T2}\end{equation}
The zeros of $L_{q-1}(k)$ are then given via 
(\ref{3a}) (taken with non positive imaginary part) 
by the zeros of the analytic continuation in energy of 
the absolute value of ${\cal T}(E)$. 
Finally, $R_+(k)$ can be obtained from (\ref{4}) and we have 
fully reconstructed the S-matrix from the Pad\'e 
approximation for the energy dependence of the 
transmittance, in the absence of bound states.

3. Although similar, (\ref{3}) differs from (\ref{1}) by 
having a {\em variable unknown} effective mass.

In $Al_cGa_{1-c}As$, $m_{eff}$ increases by about 50\% 
from the $GaAs$ value in the direct gap concentration range, 
$0\le c<0.45$.  There are two alternative ways to tackle this problem. 
One may reformulate the inverse scattering method to deal with 
equation (\ref{3}). We prefer to follow an alternative,
two-step approach. First, we solve 
an auxiliary problem: we find a potential for 
the constant mass Schr\"odinger equation (\ref{1})
such that the corresponding $S$-matrix coincides 
with the one reconstructed from the transmittance, 
as in point 2 above. 
In particular, if the scattering data are expressed by Pad\'e 
approximants, then the solution of the inverse problem is much 
simplified, and it can be accomplished by solving a linear system 
of equations \cite{KM55,Sab83a,CS89}. 
Second, we will use a unitary transformation linking constant 
and variable mass Hamiltonians \cite{BBM95} and 
derive a differential equation for the mass 
(concentration) profile in (\ref{3}). 

In fact, let $X(\xi)$ be a twice differentiable 
function, satisfying $X(\pm\infty)=\pm\infty$ and 
having a positive derivative 
$\frac{{\rm d}X}{{\rm d}\xi}=\dot X(\xi)>0.$ 
Here and in the following we use the dot to denote 
derivation with respect to $\xi$. Then,
let us consider the coordinate transformation 
\begin{equation}
x=X(\xi), \label{x}
\end{equation}
which maps the interval $(-\infty,+\infty)$ onto itself. 
As shown in \cite{BBM95}, each such change of variables
corresponds to a unitary operator, $U_X$, in the space
of square integrable functions,
such that in the new coordinate $\xi$ 
the equation (\ref{1}) transforms to 
\begin{equation}
-\frac{{\rm d}}{{\hbox{\rm d}}\xi}\frac{\hbar^2}{2m(\xi)}
\frac{{\rm d}}{{\hbox{\rm d}}\xi}f(\xi) + 
V_X(\xi)f(\xi)=Ef(\xi),\label{X1}
\end{equation}
where 
\begin{equation}
m(\xi)=m_0\dot X^2, \label{X2a}
\end{equation}
\begin{equation}
V_X(\xi)=V\left[X(\xi)\right]+\frac{\hbar^2}{8m_0\dot X^4}
\left(2\mathop{X}\limits^{\mbox{...}}\dot X-5\ddot X^2\right),\label{X2b}
\end{equation}
and where we omitted for simplicity the $\xi$ dependence of the 
derivatives of $X$. Eqs. (\ref{X1}-\ref{X2b}) and (\ref{3}) coincide if
\begin{equation}
V_X(\xi)={\cal B}\left[m(\xi)-m_\infty\right], \label{X5}
\end{equation}
where $m(\xi)$ and $V_X(\xi)$ are given by 
(\ref{X2a}, \ref{X2b}). This is a nonlinear differential 
equation of third order for $X$. Noting that it does 
not depend explicitly on $\xi$, and introducing a new 
dependent variable $S(X)=\dot X$ yields
\begin{equation}
2S^{\prime\prime}S-3S^{\prime\,2}=
\frac{8m_0S^2}{\hbar^2}\left[{\cal B}m_0(S^2-S_\infty^2)-V(X)\right], 
\label{X6}\end{equation}
The prime denotes derivation with respect to $X$. 

To be physically feasible, the solution of (\ref{X6}) 
must be regular and bounded by 
\begin{equation}
\underline{m}\le{m_0}S^2(\xi)\le\overline{m}, \label{X8}
\end{equation}
where $\underline{m}$ and $\overline{m}$ are the minimal 
and maximal achievable values of the effective mass, 
and tend  to $\sqrt{m_\infty/m_0}$ in the limits 
$X\to\pm\infty$. A necessary condition for the latter 
condition to hold is ${\cal B}>0$, which is validated  for 
$GaAlAs$. The asymptotic value of the effective mass
for the embedding crystal is a free parameter. Its value 
must be chosen in a range leading to a feasible concentration 
range, satisfying (\ref{X8}). 

In practice, we truncate the auxiliary (constant mass) potential 
$V(X)$ to a finite interval $(0,L)$. For zero potential Eq. (\ref{X6}) 
can  be explicitly solved:
$S(X)=\left[a+b{\rm e}^{\kappa X}+
c{\rm e}^{-\kappa X}\right]^{-1},$
where $\hbar^2\kappa^2=8{\cal B}m_\infty$ 
and the constants  $a,\,\,b,\,\,c$ satisfy $a^2-4bc=m_0/m_\infty$. 
Now, let $S(X)$ be a regular solution on one of the intervals 
$(-\infty,0)$ or $(L,+\infty)$. It satisfies
\begin{equation}S^\prime(X) = 
\pm\kappa S(X)\left[1-\sqrt{m_0/m_\infty}S(X)\right] 
\label{X12}\end{equation}
at all points of the respective interval, with the minus 
sign on the former and the plus on the latter. Thus, in order 
to be continuous and achievable, a solution of (\ref{X6}) on 
$(0,L)$ must satisfy (\ref{X12}) at the ends of the interval.
This is a nonlinear Sturm-Liouville problem 
that can be solved by the shooting method.
The new coordinate can be obtained then by 
integrating $\dot{X}(\xi)=S(X)$.
Reasonable precision may also be obtaineded by using a 
quasi-classical approximate solution to (\ref{X6}). 

Thus, we obtain a continuous distribution of mass (concentration) 
yielding the desired transmittance. Technological constraints 
make desirable concentration distributions in the form of a sequence of 
steps of digitized length (in lattice constants) and height 
(linear combinations of several concentrations). This can be achieved 
by approximating the continuous concentration profile and by using standard 
optimization techniques. The brute force approach to the inverse problem 
is to find the transfer matrix of (\ref{3}) for concentration 
profiles made of discrete steps (parameterized by concentrations and 
lengths) and constrain the resultant transmittance to the predetermined 
profile. This suffers from the fact that the corresponding optimization 
problem is non-convex and, even using the largest computers, satisfactory 
results are not guaranteed. To the contrary, our approach offers a good 
starting point for the optimization process. A modest exploration of its 
neighborhood usually provides superb results.

\begin{figure}[h]
\includegraphics{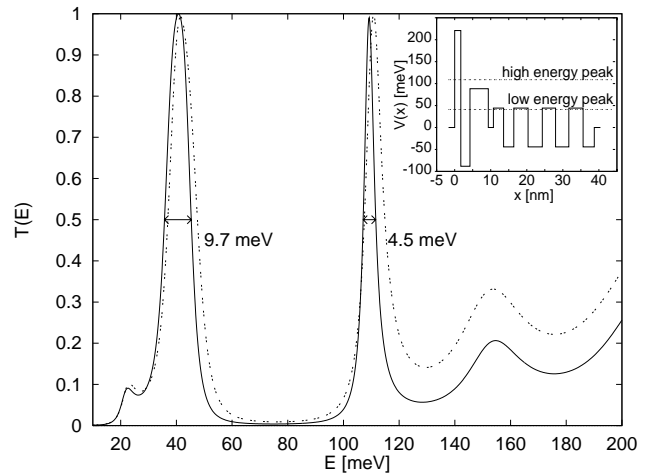}
\caption{Energy dependence of the transmittance of 12 layer digitized
$Al_cGa_{1-c}As$ filter: continuous line --- Eq.(\protect{\ref{3}});
dotted line --- constant mass approximation Eq.(\protect{\ref{1}})
with $m_0=m_\infty$. Insert: potential energy profile.\label{fig1}}
\end{figure}

As an illustration, we give in Table \ref{table1} the data for a 12 
layer $Al_cGa_{1-c}As$ filter with digitized steps in width and 
concentration.  Quite remarkably, this structure accomplishes the 
goal of having a lower energy resonance more than twice wider than 
the high-energy one. In Fig. \ref{fig1} the transmittance computed 
from (\ref{3}) is compared to a constant effective mass approximation 
using (\ref{1}) with $m_0=m_\infty$. The latter's errors become 
important at the higher resonance, whose width is overestimated by a 
factor of two, thereby losing the main feature of this structure!
The computation of the continuous profile takes several seconds on an 
IBM RISC 6000 workstation. The digitizatization and final optimization 
were done manually.

Spatial variations of the band gap have been crucial for the success 
of man-made heterostructures. We have outlined a technique which takes 
into account also the spatial variations of effective mass, and can be 
employed in the design of various electronic devices.

{\bf Acknowledgment.} 
This research supported in part by NSF and the CAU Center for 
Environmental Policy, Education and Research under 
EPA Assistance ID \# CR818689. We thank Thomas K. Gaylord and 
Elias Glytsis for bringing this problem to our attention 
together with a wealth of information on semiconductor 
heterostructure design. We thank Roger Balian, 
Khosrow Chadan and Pierre Sabatier for stimulating 
discussions on inverse scattering. 

\bibliographystyle{prsty}


\begin{table}
\caption{12 layer digitized $Al_cGa_{1-c}As$ filter.
\label{table1}}
\begin{tabular}{cccc}
Layer&Width&Width&Al\cr 
\# &(unit \tablenote{monoatomic layer width = $0.2827$ nm}) 
&(nm)
&concentration
\tablenote{$c_1=0.05714,~c_2=2c_1,~c_3=4c_1,$ bulk concentration = $c_2$}
\cr
\tableline
1&6&1.696&$c_1+c_2+c_3$\\
2&9&2.543&0\\
3&18&5.088&$c_3$\\
4&5&1.413&$c_2$\\
5&10&2.827&$c_1+c_2$\\
6&10&2.827&$c_1$\\
7&14&3.957&$c_1+c_2$\\
8&14&3.957&$c_1$\\
9&13&3.675&$c_1+c_2$\\
10&13&3.675&$c_1$\\
11&14&3.957&$c_1+c_2$\\
12&11&3.109&$c1$\\
\end{tabular}
\end{table}
\vskip2.45in

\end{document}